# Addressing Raman features of individual layers in isotopically labeled Bernal stacked bilayer graphene

**Sara D Costa, Johan Ek Weis, Otakar Frank, Michaela Fridrichová and Martin Kalbac**[*]

J. Heyrovský Institute of Physical Chemistry, Academy of Sciences of the Czech Republic, v.v.i., Dolejškova 3, CZ-18223 Prague 8, Czech Republic

E-mail: martin.kalbac@jh-inst.cas.cz

The most important bands for the evaluation of strain in graphene ( the 2D and 2D′ modes) are investigated. It is shown that for Bernal-stacked bilayers, the two-phonon Raman features have three different components that can be assigned to processes originating solely from the top graphene layer, bottom graphene layer, and from a combination of processes originating both from the top and bottom layers. The individual components of the 2D and 2D′ modes are disentangled. The reported results enable addressing the properties of individual graphene layers in isotopically labelled turbostratic and Bernal-stacked graphene systems.

**Introduction**

Employing new transfer techniques, it is possible to stack the 2-D materials in a controlled way, thus obtaining rationally designed nanostructures [1]. However, for the further development and characterization of these materials, it is crucial to address the properties of individual layers. For this purpose, Raman spectroscopy is one of the most useful spectroscopic techniques [2]. It combines simplicity, non-destructibility, and high efficiency, as it provides large amounts of information in a relatively fast way. Nevertheless, to be able to extract important information from the Raman spectra, the phonon processes must be understood and the band must be correctly assigned, which may be a challenging task, especially in multilayered graphene samples. For monolayer graphene (1-LG), three main features are observed: the G band (~1580 cm$^{-1}$), the D band (~1350 cm$^{-1}$, at 2.54 eV, only in the presence of disorder), and the 2D (or G′) band (~2700 cm$^{-1}$, at 2.54 eV). The latter comes from a two-phonon process [3], with contributions from both overtone and combination modes [4]. The 2D band for 1-LG can be usually fitted by a single Lorentzian, whereas its line shape becomes more complex for two or more graphene layers. The 2D band of bilayer graphene (2-LG) has typically been fitted with four peaks, each assigned to a different resonance process [5]. However, recent calculations have shown that Raman bands from two of the processes appear at similar frequencies, and thus, only three Lorentzian peaks are needed to fit the 2D band of 2-LG [6]. In addition to the 2D mode, other two-phonon modes appear in the Raman spectrum of graphene, such as D + D″ (~2450 cm$^{-1}$) and 2D′ (~3250 cm$^{-1}$) [7]. The D + D″ mode originates from phonons along the KΓ high-symmetry line, and its asymmetry is due to phonons from interior of the two-dimensional Brillouin zone [8]. The 2D′ mode refers to the second order of the D′ band, which is assigned to an intra-valley process [9]. Because the D + D″ and 2D′ two-phonon processes, their line shape is presumably also dependent on the number of layers and the interactions between the layers. May *et al.*[8] investigated the evolution of the D + D″ mode in dependence on the number of layers (one to three layers), and observed additional contributions for bilayer and trilayer (3-LG) graphene, in contrast to 1-LG. These contributions are assigned to additional scattering processes, originating from two or three valence and conduction bands, for 2-LG and 3-LG, respectively. Overall, it is expected that all modes originating in two-phonon processes show an evolution with the increase in the number of layers. However, even though two-phonon processes for 2-LG have been studied, the contribution of each layer is not yet completely understood. In addition, it has been recently found that phonons originating in one graphene layer can be scattered by structural disorder in the next graphene layer [10]. Isotopic labelling can be employed to address properties of individual layers because the Raman frequencies are sensitive to the different atomic masses: $\omega_{13} = \omega_{12} \times 0.963$, where $\omega_{12}$ and $\omega_{13}$ are the frequencies of the $^{12}$C and $^{13}$C Raman modes, respectively [11]. In this way, the Raman signals of the two layers appear separated in the Raman spectra, allowing identification of each layer individually [2]. In general, isotopic labelling of graphene layers has been introduced as it broadens the applicability of Raman spectroscopy in the study of multilayer graphene samples [12]. This method has been successfully applied to investigate interaction with the substrate and heating effects [13], doping by fluorination [11], and defects [10]. However, the analysis of the contribution of individual graphene layers to the Raman bands appears to be more difficult for AB-stacked bilayers. In AB-graphene, the hexagonal networks of carbon atoms are stacked perpendicularly to the layer plane, in such a way that the vacant centres of the hexagons of one layer are aligned with a carbon atom of another layer. In turbostratic graphene (T) there is no stacking order between two layers and the interlayer space is larger than for AB-stacked graphene [14]. This results in a weak interaction between adjacent layers and the Raman spectrum is similar to 1-LG, albeit all of the modes exhibit two peaks, belonging to processes either from $^{12}$C or $^{13}$C atoms. On the other hand, the Raman spectrum of AB 2-LG is more complex as it involves modes assigned to $^{12}$C, $^{13}$C, and both simultaneously. Isotopic labelling can be helpful in understanding the origin of different components of each mode. Nevertheless, the correct method for the analysis of two-phonon modes has not been established yet. The 2D band-fitting process seems to be very complex for isotopically labelled 2-LG. Araujo *et al.* [15] suggested the use of eight Lorentzian peaks, justified by a symmetry breaking in the unit cell, due to the atomic mass difference. Fang *et al.* [16] discussed how the 2D band for both $^{12}$C-bilayer graphene and $^{13}$C 2-LG can be fit with four Lorentzians. However, in $^{13}$C/$^{12}$C 2-LG the atomic mass difference can break the degeneracy of some modes, and up to eight peaks would be necessary to fit fully the 2D band. Hence, the mechanism that leads to the increase in the number of peaks in the 2D band upon isotope labelling still remains unclear. Moreover, both works mentioned above [15, 16] were based on a four-peak fitting of the 2D band, while it has recently been shown that a three-peak fitting of the 2D mode is more appropriate [6]. The components of the 2D′ mode have not been studied yet. Although the 2D′ mode intensity is relatively small, the investigation of the 2D′ band is particularly important because it represents a similar but less complex Raman mode when compared with its counterpart, the 2D band [17]. In addition, the complexity of the 2D band for 2-LG can be elucidated using the 2D′ band as a guideline. Another key advantage in studying the 2D′ band lies in the fact that its frequency is affected by strain, but much less by doping, unlike the G mode [18]. Indeed, the effects of strain on the 2D′ mode have been investigated in several works [18, 19], and the applicability of using this Raman feature for strain sensing in graphene has been confirmed. Under biaxial strain, the 2D′ mode shifts by ~110 cm$^{-1}$/% [19] without experiencing changes to its line shape, while under uniaxial strain, splitting accompanies the shift with a maximum difference between the two components of ~21 cm$^{-1}$/% [18, 19b]. In this work, 2-LG as a model system for artificially designed multi-layered systems has been studied. We also investigated the role of stacking order on the phonon interaction in the Raman active second-order processes. We analysed and clarified the origin of different components of the 2D and 2D′ bands. Moreover, the Raman contributions of each individual layer in bilayer graphene were observed using isotopic labelling, revealing properties that are not observed in regular monoisotopic bilayer graphene. In particular, based on the observation of the 2D′ mode for labelled 2-LG in the Bernal configuration, it was found that the 2D band also comprises modes originating from both layers simultaneously.

**Materials and Methods**

The graphene sample preparation has been described elsewhere [11]. Briefly, the graphene was synthesized by chemical vapour deposition (CVD). A copper foil was heated to 1273 K and annealed for 20 min under a flow of 50 sccm H$_2$. 1 sccm of methane ($^{12}$CH$_4$) was then introduced to the chamber for 90 s, which resulted in the growth of graphene containing $^{12}$C.

The $^{12}CH_4$ was turned off and a flow of 1 sccm $^{13}CH_4$ was introduced for 30 min. The sample was cooled to room temperature. The graphene was subsequently transferred to a $SiO_2$/Si substrate, as reported previously [20]. In brief, the substrate was covered by poly(methyl methacrylate) (PMMA). The copper was subsequently etched away and the PMMA-covered graphene was transferred to a $SiO_2$/Si substrate. Residual PMMA was removed by heating the sample to 773 K in a hydrogen and argon atmosphere. Raman spectra were acquired at room temperature using a LabRAM HR Raman spectrometer (Horiba Jobin-Yvon), with an integrated He–Ne laser of 1.96 eV. The samples were also excited using an extra laser source of an Ar/Kr laser (2.54, 2.41, 2.34, and 2.19 eV, Coherent). A 50× objective was used, providing a laser spot of about 1 μm. The laser power was kept below 1 mW to avoid overheating the sample.

**Results and discussion**

The frequencies of the Raman features are sensitive to changes in the atomic mass. Therefore, for 2-LG with one layer composed of $^{12}C$ carbon and the other of $^{13}C$ carbon, the Raman modes should be separated for the different layers. The Raman spectra for T and AB-stacked 2-LG are shown in figure 1, in which the 'graphitic mode' (G band), the 'defect mode' (D band), and three two-phonon modes, 2D, D + D″, and 2D′ bands, are depicted.

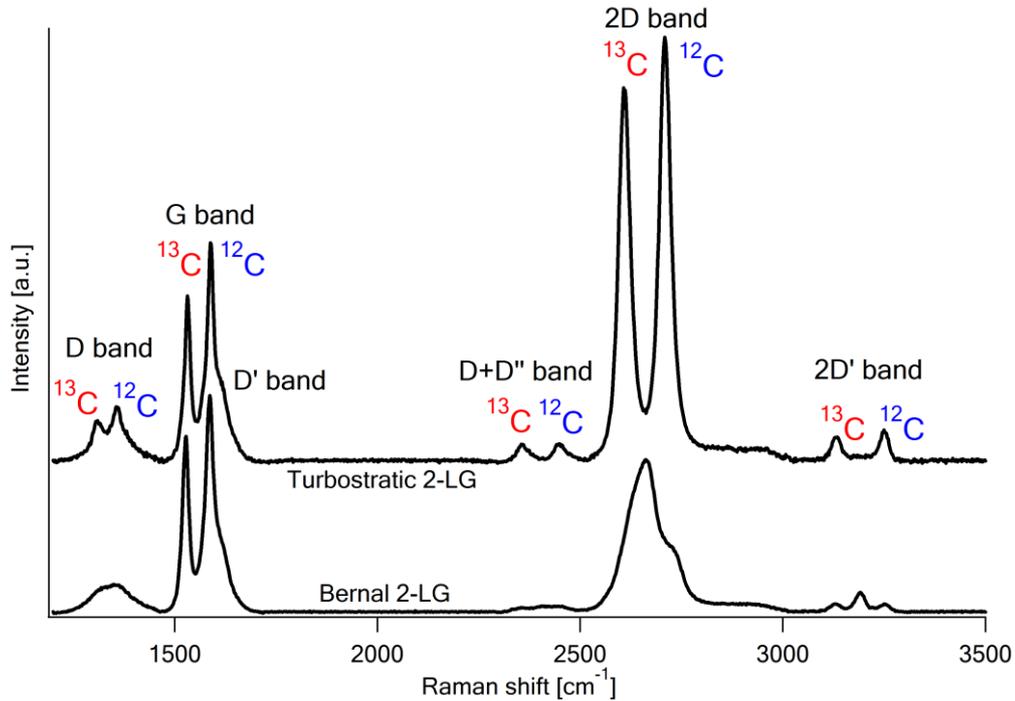

Figure 1. Raman spectra of $^{13}C/^{12}C$ bilayer graphene, at 2.54 eV: with Bernal stacking, AB 2-LG (bottom spectrum) and turbostratic, T 2-LG (top spectrum). The spectra show the main Raman modes: D, G, D′, 2D, D + D″, and 2D′. The spectra were normalized to the intensity of the G band.

In general, for turbostratic 2-LG, the bands stemming from two-phonon processes exhibit sharper line shapes, while for AB 2-LG the two-phonon process bands are broader. This effect is especially well pronounced in the case of the 2D band. For T 2-LG, the 2D band is represented by two relatively sharp lines, assigned to the bottom and top graphene layers (and to the two different isotopes), while for AB 2-LG, only one broad band corresponding to the 2D mode is present in the Raman spectrum. In the latter case, both the top and bottom graphene layers contribute to the broad 2D and there is no clear separation between the contributions of each layer. The differences in the 2D band originate from the interaction between the layers, which is weak in a turbostratic configuration and relatively strong for AB-stacked layers [21]. In this way, the spectrum for turbostratic 2-LG (of natural isotope content) ends up with similar characteristics to one of 1-LG, which shows a single and symmetric 2D band.

Most of the time, the two kinds of graphene can be distinguished by comparing their 2D full width at half maximum (FWHM) values: ~25 cm$^{-1}$ for 1-LG and up to ~50 cm$^{-1}$ for T 2-LG [3, 14, 16, 22]. For isotopically modified turbostratic 2-LG, the FWHM is ~37 cm$^{-1}$ for both $^{13}$C and $^{12}$C components. This indicates that the higher value of FWHM found for $^{12}$C/$^{12}$C 2-LG is due to the contributions from the top and bottom layers being slightly shifted because of a difference in strain, doping, or to a small mismatch in stacking angle [22], resulting in a broader peak.

Another interesting feature in the Raman spectra of graphene is the 2D′ band, found in the range 3120–3250 cm$^{-1}$. For turbostratic 2-LG, the 2D′ band splits into two modes, assigned to the different isotopes, but for Bernal-stacked 2-LG, three modes are observed. The central mode, which does not appear in the Raman spectrum of turbostratic 2-LG, is assigned to a phonon process involving contributions from both $^{13}$C and $^{12}$C layers. Figure 2a shows the 2D′ band in detail. The 2D′ band is an intra-valley two-phonon process, and the two phonons can either originate from the same layer or one phonon from each graphene layer. Consequently, three components of the 2D′ mode are expected for isotopically labelled AB-stacked bilayer graphene: 3130 cm$^{-1}$ for 100% $^{13}$C (two phonons from the $^{13}$C layer), 3190 cm$^{-1}$ for 50% $^{13}$C (one phonon from the $^{13}$C layer and one from the $^{12}$C layer), and 3250 cm$^{-1}$ for 0% $^{13}$C or pure $^{12}$C (two phonons from the $^{12}$C layer). The area of the mode found at 3190 cm$^{-1}$ is twice as large as the area of the other two components, as it comprises both the $^{12}$C → $^{13}$C and the $^{13}$C → $^{12}$C components. For turbostratic graphene, the 2D′ band shows only two components, at 3130 cm$^{-1}$ for $^{13}$C and at 3250 cm$^{-1}$ for $^{12}$C graphene. This is again justified by the weak coupling between the layers, as there is no stacking order [14]. These observations corroborate the general hypothesis that for isotopically labelled AB 2-LG, the phonons of the top and bottom layers can combine, resulting in additional Raman modes. A similar effect is expected for the D + D″ band, at around 2450 cm$^{-1}$. However, because of the asymmetry of this band and the complexity of its origin [4, 8], its shape is not as well defined as that of the 2D′ band. The isotopic effect is still clear, because the band becomes broader.

Having the 2D′ band with the Raman modes of the top and bottom layers clearly separated by its frequencies may reveal the effects of strain on the individual layers. Evaluating the frequency of the 2D′ band for the top layer, $^{12}$C, one can say that almost no strain is observed, while for the bottom layer, compressive strain of about − 0.1% (according to reference [19a]) is observed. In Bernal-stacked bilayer graphene, it is known that both layers strongly interact with each other, usually sharing defects, doping, and strain effects. Thus, it is likely that the bottom and top layers present similar amounts of strain. Note that the variation in the frequency shift observed in the 2D′ band can be caused by a different factor, such as the stacking angle [22] which is indeed observed experimentally.

Figure 2b shows the dependence of the 2D′ band frequency on the laser energy. The measured sets of data correspond to the different components of the 2D′ mode for turbostratic and Bernal-stacked 2-LG. The dispersion slopes of all components of the 2D′ mode are similar, with values around 20 cm$^{-1}$/eV (see table 1). The dispersion slope of the 2D′ band is, thus, twice as large as for the D′ band [17], confirming the assignment of the 2D′ band to a second-order process of the D′ mode [14].

Table 1. Fitted slopes for the dispersion of the 2D′ band modes for Bernal-stacked and turbostratic graphene.

| | $\frac{\Delta\omega_2 D'}{\Delta E_l} \sim$ cm$^{-1}$eV$^{-1}$ |
|---|---|
| Bernal | 17 ± 4 |
| | 20 ± 3 |
| | 22 ± 6 |
| Turbostratic | 21 ± 2 |
| | 21 ± 7 |

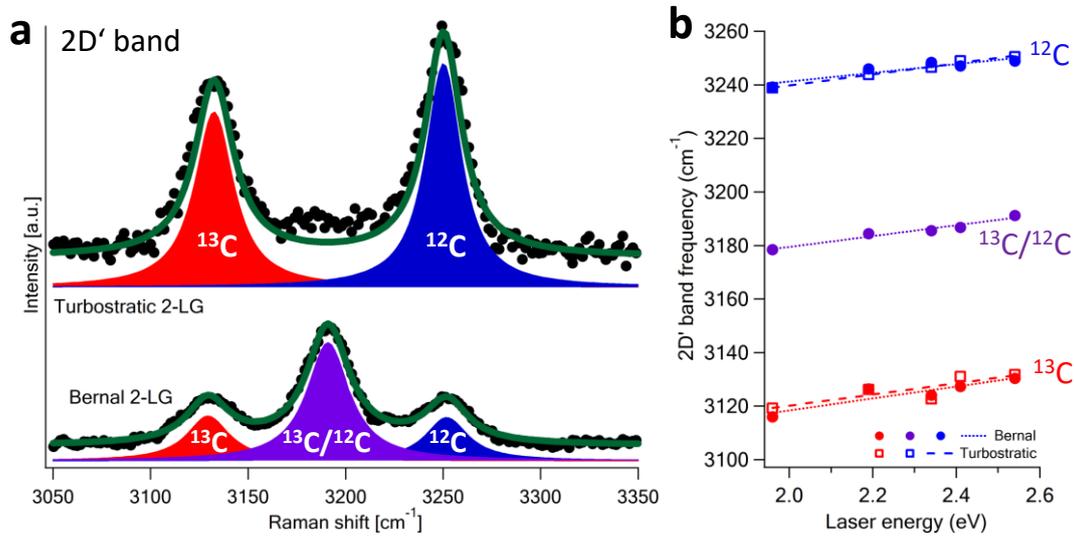

Figure 2. (a) Raman spectra in the 2D′ band region for turbostratic and Bernal-stacked 2-LG, at 2.54 eV. (b) 2D′ band dispersion for Bernal-stacked and turbostratic graphene. The corresponding linear fits are also represented, and their slopes are shown in table 1.

The 2D′ band can also be used as a guideline to the analysis of other two-phonon Raman modes, such as the 2D band. Both the 2D and 2D′ bands originate in the combination of four virtual processes of absorption of a photon, creating an electron–hole pair, followed by two consecutive processes of scattering by a phonon and the final recombination of the electron–hole pair with the emission of a photon [4]. For the 2D band, this scattering process occurs between two non-equivalent K points in the Brillouin zone of graphene (inter-valley scattering), while for the 2D′ band the scattering process occurs within the same Dirac cone (intra-valley scattering).

For bilayer graphene with Bernal stacking, the electron dispersion in the valence and conduction bands splits into two branches near the K point, and optical transitions between the two bands are allowed [17]. This splitting results in four possible inter-valley processes, as shown in the scheme presented in figure 3a: P11, P22, P12, and P21.

Upon introduction of different isotopes, the number of components of the 2D band doubles, because phonons originating in each layer will be distinguished. In other words, eight processes can emerge, four assigned to $^{12}$C and four assigned to $^{13}$C, with frequencies shifted from each other. In addition, and similarly to the 2D′ band, for AB 2-LG, processes involving two phonons originating from each form of the different layers should also be taken into account. Therefore, another set of eight modes should be considered, four assigned to $^{13}$C → $^{12}$C modes and four assigned to $^{12}$C → $^{13}$C. However, the modes involving both isotopes are degenerate, as the frequency of the modes $^{12}$C → $^{13}$C is equal to the frequency of the modes $^{13}$C → $^{12}$C, giving rise to a total of 12 components contributing to the 2D band. In addition, it has recently been found that for monoisotopic 2-LG, two of the modes are also located close together, and experimentally the 2D band is better fitted with three rather than four components [6]. Accordingly, for isotopically labelled Bernal-stacked 2-LG, the 2D band is composed of nine modes.

Figure 3b shows simplified schemes of all the possible processes that contribute to the 2D band of isotopically labelled 2-LG. The blue and red arrows are assigned to processes involving only $^{12}$C or only $^{13}$C phonons, respectively. The blue/red arrows represent processes that start in a $^{12}$C graphene layer and go to a $^{13}$C graphene layer, and the opposite for red/blue arrows. The frequency of the modes comprising phonons from both $^{12}$C and $^{13}$C layers is expected to be exactly in between the frequencies of the bands composed of phonons solely originating in the $^{12}$C layer or in the $^{13}$C layer. Figure 3c shows the fit of the 2D band using nine Lorentzians. The condition for fitting procedure was set that the sums of the bands from a particular isotope (identified by different colours) are equidistant. The frequency difference of the $^{12}$C, $^{13}$C, and $^{13}$C/$^{12}$C modes was estimated accounting for the proportion of $^{13}$C involved. Small deviations between the experimental and estimated frequencies of the components of 2D band are expected because of strain variations throughout the sample [23]. Similarly, as in the case of the 2D′ band, the intensity relation of the 2D-mode components is expected to be similar for the $^{12}$C and $^{13}$C phonons, while the intensity for the central band should double, as it comprises both processes from $^{12}$C → $^{13}$C and $^{13}$C → $^{12}$C.

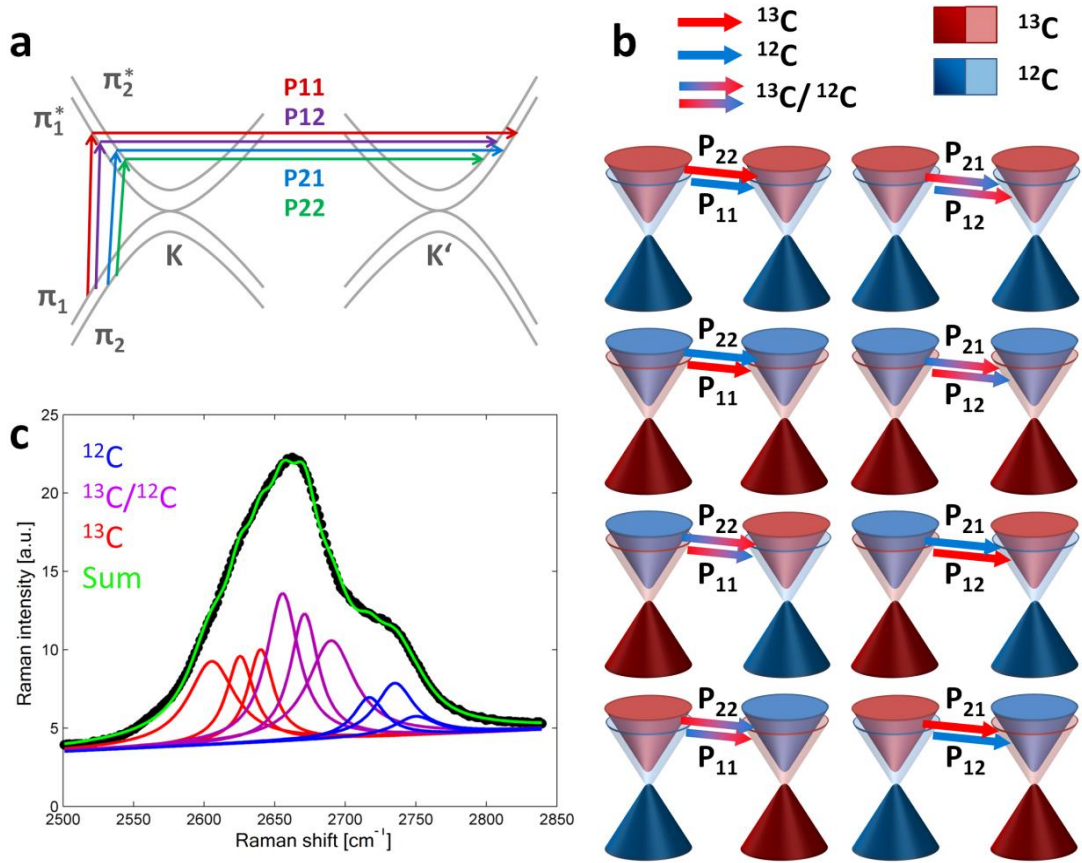

Figure 3. (a) Schematic view of the electron dispersion of AB 2-LG near the K and K′ points, showing the doubled valence ($\pi_1$, $\pi_2$) and conduction ($\pi_1$*, $\pi_2$*) bands. Four two-phonon processes are described: P11, P12, P21, and P22. (b) Schematic view of the electron dispersion of isotopically labelled AB 2-LG near the K and K′ points, showing each possible process present in the 2D band. The blue cones are assigned to $^{12}$C, while the red cones represent $^{13}$C. (c) 2D band fit for isotopically labelled AB 2-LG. The fit was performed using nine Lorentzian lines: three lines are assigned to $^{13}$C

modes (red), three are assigned to $^{12}$C (blue) modes, and the other three are assigned to modes involving both isotopes (purple).

Recently, the appearance of an additional feature, at around 2655 cm$^{-1}$, has been reported for turbostratic graphene [24]. The frequency of this peak falls exactly in between the frequencies of the $^{13}$C and $^{12}$C contributions of the 2D band, suggesting that it can originate from phonons of different graphene layers. Even though the spectrum refers to a turbostratic grain, it was found that if the twisting angle is low, i.e. the structure is close to AB stacking, the 2D band can be similar to that found for Bernal structures [22]. Consequently, in special cases, the component comprising a phonon from each layer in a bilayer structure can be observed even for turbostratic 2-LG samples.

**Conclusion**

The Raman spectra of isotopically labelled AB-stacked and T bilayers have been investigated. The application of different isotopes varies the relative mass of each layer, so different Raman signals were obtained and the layers could be addressed individually. We focused on the 2D and 2D′ bands of bilayer graphene, which are crucial for the evaluation of mechanical strain in graphene. It was found that the 2D′ band is composed of two modes in turbostratic 2-LG, one for each layer, while for AB 2-LG three modes were observed, assigned to $^{13}$C, $^{13}$C/$^{12}$C, and $^{12}$C processes. The third mode found for AB-stacked layers was assigned to scattering processes involving both isotopes. For the 2D band of labelled bilayer graphene, 16 processes have been identified, although only nine modes are distinguished experimentally. Each set of three bands can be attributed to processes originating either in one layer or to a combination of processes from the two layers. Thus, the properties of phonons in graphene bilayers were addressed and we believe that our conclusions can be extended to other multilayered structures or even to systems based on different 2-D materials.

**Acknowledgements**

The authors acknowledge the support of MSMT ERC-CZ project (LL 1301).